\documentclass{iaus}
\usepackage{graphicx}

\title[Binary central stars] 
{Ongoing surveys for close binary central stars and wider implications}

\author[Brent Miszalski]   
{Brent Miszalski}

\affiliation{
$^1$South African Astronomical Observatory and Southern African Large Telescope Foundation, PO Box 9, Observatory, 7935, South Africa\\[\affilskip]
email: {\tt brent@saao.ac.za}
}

\pubyear{2011}
\volume{283}  
\pagerange{1-4}
\jname{Planetary Nebulae an Eye to the Future}
\editors{A.C. Editor, B.D. Editor \& C.E. Editor, eds.}
\begin{document}

\maketitle

\begin{abstract}
   Binary central stars have long been invoked to explain the vexing shapes of planetary nebulae (PNe) despite there being scant direct evidence to support this hypothesis. Modern large-scale surveys and improved observing strategies have allowed us to significantly boost the number of known close binary central stars and estimate at least 20\% of PNe have close binary nuclei that passed through a common-envelope (CE) phase. The larger sample of post-CE nebulae appears to have a high proportion of bipolar nebulae, low-ionisation structures (especially in SN1987A-like rings) and polar outflows or jets. These trends are guiding our target selection in ongoing multi-epoch spectroscopic and photometric surveys for new binaries. Multiple new discoveries are being uncovered that further strengthen the connection between post-CE trends and close binaries. These ongoing surveys also have wider implications for understanding CE evolution, low-ionisation structure and jet formation, spectral classification of central stars, asymptotic giant branch (AGB) nucleosynthesis and dust obscuration events in PNe.
\end{abstract}

\firstsection 
\firstsection 
              
\section{Introduction}
The shapes of PNe are well known to demonstrate an amazing variety shapes (e.g. Sahai et al. 2011). Balick \& Frank (2002) reviewed many possible shaping mechanisms and in the case of binarity remarked: 
\begin{quote}
   \emph{The bottom line is that although binarity is a popular mechanism for forming axisymmetric structures on PNe, \textbf{direct evidence} to support the efficacy of the process is not strong.}
\end{quote}
At the time only $\sim$12 close binary central stars were known (Bond 2000) and no strong trends in nebula morphology were apparent (Bond \& Livio 1990). 

\begin{figure}
   \begin{center}
      \includegraphics[scale=0.275]{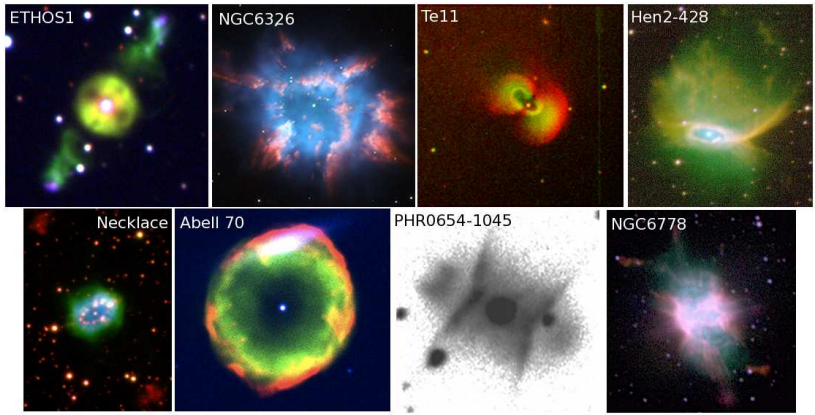}
   \end{center}
   \caption{A selection of PNe with binary central stars. Images were taken from references in the text except for Te~11 (Jacoby et al. 2010), Hen~2-428 (Manchado et al. 1996) and NGC~6778 (Miranda et al. 2010). Note the presence of jets (e.g. ETHOS~1 and Necklace), low-ionisation structures (e.g. NGC~6326 and NGC~6778) and bipolar nebulae (e.g. Hen~2-428 and NGC~6778).}
   \label{fig:fig1}
\end{figure}

Since then dramatic progress has been made with more than 40 close binaries now known (e.g. Miszalski et al. 2011a). This improvement was largely driven by the improved multiplex advantage offered by the OGLE microlensing survey (Udalski 2009) which allowed for many discoveries to be made by Miszalski et al. (2008, 2009a). Substantial progress has also been helped by improved observing strategies including targeting PNe with suspected post-CE morphologies (Miszalski et al. 2011a) and the use of less nebula-contaminated filters to obtain lightcurves (Miszalski et al. 2011d). In addition to the OGLE discoveries more close binaries were found by Hillwig et al. (2010), Hajduk et al. (2010), Corradi et al. (2011), Santander-Garc\'ia et al. (2011) and Miszalski et al. (2011a, 2011c, 2011d). Fewer wide binary systems are known and they mostly have chemically peculiar giants or sub-giants (e.g. Bond et al. 2003; Frew et al. 2011; Miszalski et al. 2011e). Figure \ref{fig:fig1} shows a selection of these new binaries. 

\firstsection 
\section{Common-envelope evolution}
The majority of binary central stars discovered have orbital periods less than 1 day (Miszalski et al. 2009a, 2011a). This is firm evidence that these systems passed through the poorly understood CE phase (Iben \& Livio 1993). Post-CE PNe offer one of the best opportunities to measure the orbital period distribution directly after the CE phase since the nebula guarantees the period is `fresh' and has not yet been altered by further angular momentum loss. This is a crucial ingredient for improving CE population synthesis models that currently rely upon a parametric formalism (see Boffin \& Miszalski 2011 for further discussion). 

With the excellent sensitivity of the OGLE survey Miszalski et al. (2009a) determined the close binary fraction in PNe to be at least $17\pm5$\% and ruled out the presence of a significant population of binaries with periods greater than $\sim$1 day. The most promising aspect of the new discoveries will be the ability to obtain accurate Keplerian masses once radial velocity orbits are measured. This will greatly improve our understanding of PN evolution that is currently heavily dependent on models for central star masses. 

\firstsection 
\section{Nebula morphologies, low-ionisation structures and jets}
Miszalski et al. (2009b) analysed images of an enlarged sample of 30 post-CE PNe to find at least 1/3 had canonical bipolar morphologies. The actual bipolar fraction may be as high as 60--70\% because of inclination effects, but this remains to be proven via kinematic modelling. A large proportion of the sample also showed low-ionisation structures (LIS, Gon{\c c}alves et al. 2001) and collimated polar outflows or jets. The LIS occur in small knots (e.g. DS~1), rings of knots (e.g. the Necklace) or in more extreme configurations (e.g. NGC~6326) and may be produced by a photoionising wind interacting with dust and gas placed in the orbital plane during the CE phase. If this scenario is correct, then the uncanny resemblance between SN1987A and the Necklace may imply a binary progenitor was responsible for the bipolar nebula and ring of knots seen in SN1987A (see e.g. Boffin \& Miszalski 2011). The extreme set of knots in NGC~6326 also suggest a binary origin for morphologically similar features in A~30 (Miszalski et al. 2011d).

Jets are prominent in A~63, the Necklace and ETHOS~1 providing the first clear evidence for a binary origin of these features (Soker \& Livio 1994). Kinematics of these nebulae show the jets are older than the main nebula, suggesting that they were launched by a temporary accretion disk around the companion established during or prior to the CE phase. Such a configuration also explains why the jets in post-CE PNe are not highly collimated due to orbital motion and precession (see e.g. Raga et al. 2009). Some of these systems were successfully discovered by pre-selecting for these nebula trends (Miszalski et al. 2011a), however a much larger campaign is needed to verify these initial results.

\firstsection 
\section{The ``wels'' are not a stellar classification}
It has long been known that close binary central stars, consisting of a white dwarf primary and a main sequence secondary, display narrow emission lines of C~III, N~III and C~IV originating near an irradiated zone on the secondary (e.g. Pollacco \& Bell 1994). Miszalski et al. (2011c, 2011d) emphasised the real overlap between the orbit-dependent irradiated emission and the so-called ``weak emission line'' central stars (\emph{wels}, Tylenda et al. 1993). Many \emph{wels} could therefore turn out to be close binary stars if high quality time-series spectroscopy were to reveal radial velocity shifts in these lines and confirmed by periodic lightcurves. This would go a large way towards explaining the \emph{wels} which do not fit any central star classification schemes (M\'endez 1991; Crowther et al. 1998). Those which are not found to be close binaries might be classified as [WN/WC] (Todt et al. 2010) or [WC]-PG1159 (Parthasarathy et al. 1998). 

In summary, the \emph{wels} are a description of spectroscopic features and not a stellar classification to be used in central star catalogues. In close binaries the weak emission lines do not originate from the primary, as proven by NGC~6326 (Miszalski et al. 2011d), and therefore cannot be considered for the basis for any classification. If not a close binary, then the object should be left as unclassified or assigned into the [WN/WC] or [WC]-PG1159 class if there is sufficient supporting information. Analysis of new examples befitting these latter classes will be presented elsewhere.

\firstsection 
\section{AGB nucleosynthesis, Type-I PNe and the s-process}
In the sample of known close binaries only NGC~6778 (Miszalski et al. 2011d) shows Type-I chemical abundances (Perinotto et al. 2004). The high He/H and N/O ratios seen have long been considered to be only produced by progenitors massive enough to experience hot bottom burning ($M>4$ $M_\odot$), however in this case we may be seeing an alternative close binary channel to produce Type-I PNe (Karakas et al. 2009; Karakas \& Lugaro 2010). Keplerian masses of NGC~6778 and future expected discoveries will shed much greater insight into the origin of Type-I PNe than hitherto possible. 

The level of s-process enrichment in nebulae (Sterling \& Dinerstein 2008) and Barium central stars (e.g. Bond et al. 2003; Miszalski et al. 2011e) is also a valuable tool to further understand Type-I PNe (see also Karakas \& Lugaro, these proceedings). Work so far suggests that binarity or some other process may be reducing the s-process enrichment in Type-I PNe, although the binary fraction of the Sterling \& Dinerstein (2008) sample is yet to be determined. Abell 70 is a particularly interesting case since it is a s-process enhanced G8IV-V companion and shows Type-I abundances at $\sim$2 kpc below the Galactic Plane, far away from where massive progenitors are expected (Miszalski et al. 2011e).

\section{Dust obscuration events and binarity}
The central stars of three peculiar PNe are known to show R Coronae Borealis-like dust obscuration events: NGC~2346 (M\'endez et al. 1982), Hen~3-1333 (Cohen et al. 2002) and M~2-29 (Hajduk et al. 2008). Miszalski et al. (2011b) showed that the quasi-periodic events are not triggered by a binary companion interacting at periastron, but rather occur at the inner edge of a dust disk at $\sim$100 AU (e.g. Chesneau et al. 2006). Such dust disks are however thought to depend upon binaries for their formation (Van Winckel et al. 2009; Chesneau 2011). So far only NGC~2346 has been proven to be a binary. Miszalski et al. (2011b) proved that the Galactic Bulge PN M~2-29 does not have a close binary central star as previously claimed and suggested instead that an unobserved yellow subgiant may be present to account for the D'-type symbiotic star features. 

\begin{acknowledgements}
I would like to thank all my collaborators for contributing to the rapid rise of binary central stars now known in PNe. 

\end{acknowledgements}

\firstsection 


\begin{thebibliography}{}
  \bibitem[Balick \& Frank(2002)]{2002ARAA..40..439B} Balick, B., \& Frank, A.\ 2002, ARA\&A, 40, 439 
\bibitem[Boffin \& Miszalski(2011)]{2011arXiv1108.0389B} Boffin, H.~M.~J., \& Miszalski, B.\ 2011, arXiv:1108.0389 
\bibitem[Bond \& Livio(1990)]{1990ApJ355..568B} Bond, H.~E., \& Livio, M.\ 1990, ApJ, 355, 568 
\bibitem[Bond(2000)]{2000ASPC..199..115B} Bond, H.~E.\ 2000, Asymmetrical Planetary Nebulae II, 199, 115 
\bibitem[Bond et al.(2003)]{2003AJ....125..260B} Bond, H.~E., Pollacco, D.~L., \& Webbink, R.~F.\ 2003, AJ, 125, 260 
\bibitem[Chesneau et al.(2006)]{2006A&A...455.1009C} Chesneau, O., et al.\ 2006, A\&A, 455, 1009 
\bibitem[Chesneau(2011)]{2010arXiv1010.1081C} Chesneau, O.\ 2011, Asymmetric Planetary Nebulae V, arXiv:1010.1081 
\bibitem[Cohen et al.(2002)]{2002MNRAS.332..879C} Cohen, M., Barlow, M.~J., Liu, X.-W., \& Jones, A.~F.\ 2002, MNRAS, 332, 879 
\bibitem[Corradi et al.(2011)]{2011MNRAS.410.1349C} Corradi, R.~L.~M., et al.\ 2011, MNRAS, 410, 1349 
\bibitem[Crowther et al.(1998)]{1998MNRAS.296..367C} Crowther, P.~A., De Marco, O., \& Barlow, M.~J.\ 1998, MNRAS, 296, 367 
\bibitem[Frew et al.(2011)]{2011PASA...28...83F} Frew, D.~J., et al.\ 2011, PASA, 28, 83 
\bibitem[Gon{\c c}alves et al.(2001)]{2001ApJ...547..302G} Gon{\c c}alves, D.~R., Corradi, R.~L.~M., \& Mampaso, A.\ 2001, ApJ, 547, 302 
\bibitem[Hillwig et al.(2010)]{2010AJ....140..319H} Hillwig, T.~C., Bond, H.~E., Af{\c s}ar, M., \& De Marco, O.\ 2010, AJ, 140, 319 
\bibitem[Hajduk et al.(2008)]{2008A&A...490L...7H} Hajduk, M., Zijlstra, A.~A., \& Gesicki, K.\ 2008, A\&A, 490, L7      
\bibitem[Hajduk et al.(2010)]{2010MNRAS.406..626H} Hajduk, M., Zijlstra, A.~A., \& Gesicki, K.\ 2010, MNRAS, 406, 626 
\bibitem[Iben \& Livio(1993)]{1993PASP..105.1373I} Iben, I.~J., \& Livio, M.\ 1993, PASP, 105, 1373 
\bibitem[Jacoby et al.(2010)]{2010PASA...27..156J} Jacoby, G.~H., Kronberger, M., Patchick, D., et al.\ 2010, PASA, 27, 156 
\bibitem[Karakas et al.(2009)]{2009ApJ...690.1130K} Karakas, A.~I., et al.\ 2009, ApJ, 690, 1130 
\bibitem[Karakas \& Lugaro(2010)]{2010PASA...27..227K} Karakas, A.~I., \& Lugaro, M.\ 2010, PASA, 27, 227 
\bibitem[Karakas \& Lugaro(2011)]{2011IAUS} Karakas, A.~I., \& Lugaro, M.\ 2011, IAUS, 283
\bibitem[Manchado et al.(1996)]{1996iacm.book.....M} Manchado, A., et al.\ 1996, The IAC morphological catalog of northern Galactic PNe
\bibitem[M{\'e}ndez et al.(1982)]{1982A&A...116L...5M} M{\'e}ndez, R.~H., Gathier, R., \& Niemela, V.~S.\ 1982, A\&A, 116, L5 
\bibitem[M{\'e}ndez(1991)]{1991IAUS..145..375M} M{\'e}ndez, R.~H.\ 1991, Evolution of Stars: the Photospheric Abundance Connection, 145, 375 
\bibitem[Miranda et al.(2010)]{2010PASA...27..180M} Miranda, L.~F., Ramos-Larios, G., \& Guerrero, M.~A.\ 2010, PASA, 27, 180 
\bibitem[Miszalski et al.(2008)]{2008A&A...488L..79M} Miszalski, B., Acker, A., Moffat, A.~F.~J., Parker, Q.~A., \& Udalski, A.\ 2008, A\&A, 488, L79 
\bibitem[Miszalski et al.(2009)]{2009A&A...496..813M} Miszalski, B., Acker, A., Moffat, A.~F.~J., Parker, Q.~A., \& Udalski, A.\ 2009a, A\&A, 496, 813 
\bibitem[Miszalski et al.(2009)]{2009A&A...505..249M} Miszalski, B., Acker, A., Parker, Q.~A., \& Moffat, A.~F.~J.\ 2009b, A\&A, 505, 249 
      \bibitem[Miszalski et al.(2011)]{2011apn5.confE.328M} Miszalski, B., et al.\ 2011a, Asymmetric Planetary Nebulae V, arXiv:1009.2890 
   \bibitem[Miszalski et al.(2011)]{2011A&A...528A..39M} Miszalski, B., et al.\ 2011b, A\&A, 528, A39 
\bibitem[Miszalski et al.(2011)]{2011MNRAS.413.1264M} Miszalski, B., et al.\ 2011c, MNRAS, 413, 1264 
   \bibitem[Miszalski et al.(2011)]{2011A&A...531A.158M} Miszalski, B., et al.\ 2011d, A\&A, 531, A158 
   \bibitem[Miszalski et al.(2011)]{2011arXiv1108.3957M} Miszalski, B., et al.\ 2011e, MNRAS, in press, arXiv:1108.3957 
\bibitem[Parthasarathy et al.(1998)]{1998A&A...329L...9P} Parthasarathy, M., Acker, A., \& Stenholm, B.\ 1998, A\&A, 329, L9 
\bibitem[Perinotto et al.(2004)]{2004MNRAS.349..793P} Perinotto, M., Morbidelli, L., \& Scatarzi, A.\ 2004, MNRAS, 349, 793 
\bibitem[Pollacco \& Bell(1994)]{1994MNRAS.267..452P} Pollacco, D.~L., \& Bell, S.~A.\ 1994, MNRAS, 267, 452 
\bibitem[Raga et al.(2009)]{2009ApJ...707L...6R} Raga, A.~C., Esquivel, A., Vel{\'a}zquez, P.~F., et al.\ 2009, ApJL, 707, L6 
\bibitem[Sahai et al.(2011)]{2011AJ....141..134S} Sahai, R., Morris, M.~R., \& Villar, G.~G.\ 2011, AJ, 141, 134 
\bibitem[Santander-Garcia et al.(2011)]{2011apn5.confE.259S} Santander-Garcia, M., et al.\ 2011, Asymmetric Planetary Nebulae V, arXiv:1009.3055
\bibitem[Soker \& Livio(1994)]{1994ApJ...421..219S} Soker, N., \& Livio, M.\ 1994, ApJ, 421, 219 
   \bibitem[Sterling \& Dinerstein(2008)]{2008ApJS..174..158S} Sterling, N.~C., \& Dinerstein, H.~L.\ 2008, ApJS, 174, 158 
\bibitem[Todt et al.(2010)]{2010A&A...515A..83T} Todt, H., Pe{\~n}a, M., Hamann, W.-R., \& Gr{\"a}fener, G.\ 2010, A\&A, 515, A83 
\bibitem[Tylenda et al.(1993)]{1993A&AS..102..595T} Tylenda, R., Acker, A., \& Stenholm, B.\ 1993,  A\&AS, 102, 595 
\bibitem[Udalski(2009)]{2009ASPC..403..110U} Udalski, A.\ 2009, Astronomical Society of the Pacific Conference Series, 403, 110 
\bibitem[Van Winckel et al.(2009)]{2009A&A...505.1221V} Van Winckel, H., et al.\ 2009, A\&A, 505, 1221 
\end{thebibliography}
\end{document}